\documentclass[12pt]{article}
\usepackage{amsmath,amssymb}
\usepackage{graphicx}
\usepackage{cite,fancyhdr}
\pdfoutput=1

\usepackage[affil-it]{authblk}
\usepackage[left=2cm,right=2cm,top=2cm,bottom=2cm,a4paper]{geometry}

\newcommand{\vsp}{\vspace{10pt}}

\allowdisplaybreaks

\begin{document}

\setcounter{footnote}{0}
\setcounter{figure}{0}
\setcounter{table}{0}

\title{\bf \Large 
Splitting Mass Spectra and Muon $g-2$ in Higgs-Anomaly Mediation
}
\author{{\normalsize Wen Yin}}
\author{{\normalsize Norimi Yokozaki}}

\affil{\small 
Department of Physics, Tohoku University, \authorcr {\it Sendai, Miyagi 980-8578, Japan}}

\date{}

\maketitle

\thispagestyle{fancy}
\rhead{TU-1030}
\cfoot{\thepage}
\renewcommand{\headrulewidth}{0pt}

\begin{abstract}
\noindent
We propose a scenario where only the Higgs multiplets have direct couplings to a supersymmetry (SUSY) breaking sector.
The standard model matter multiplets as well as the gauge multiples are sequestered from the SUSY breaking sector;
therefore, their masses arise via anomaly mediation at the high energy scale with a gravitino mass of $\sim100$\,TeV.
%
Due to renormalization group running effects from the Higgs soft masses,
the masses of the third generation sfermions become $\mathcal{O}$(10)\,TeV at the low energy scale, 
while the first and second generation sfermion masses are $\mathcal{O}$(0.1\,-\,1)\,TeV, avoiding the tachyonic slepton problem and flavor changing neutral current problem. With the splitting mass spectrum, the muon $g-2$ anomaly is explained consistently with the observed Higgs boson mass of 125\,GeV.
Moreover, the third generation Yukawa couplings are expected to be unified in some regions.
\end{abstract}

\clearpage

\section{Introduction}
The low-energy supersymmetry (SUSY) is one of the leading candidates for the physics beyond the standard model (SM), 
and provides attractive features. The Higgs potential is stabilized against quadratic divergences. 
Three SM gauge couplings unify at around $10^{16}$\,GeV in the minimal extension, so-called the minimal supersymmetric standard model (MSSM). 
This fact suggests the existence of the grand unified theory (GUT), leading to a natural explanation of the charge quantization. 

With a discovery of the Higgs boson with a mass of 125\,GeV~\cite{Aad:2015zhl}, it tours out that 
a rather large radiative correction from scalar tops (stops) to the Higgs boson mass is required~\cite{Okada:1990vk, Okada:1990gg, Ellis:1990nz, Ellis:1991zd, Haber:1990aw}, 
since its mass is predicted to be smaller than $Z$ boson mass at the tree level in the MSSM. In the absence of a larger trilinear coupling of the stops, the stop mass is expected to be as large as $\mathcal{O}(10)$\,TeV. 
This is not very encouraging since it seems difficult to be consistent with another important motivation for the low-energy SUSY: 
the observed anomaly of the muon anomalous magnetic moment (muon $g-2$).

The muon $g-2$, $a_\mu$, is measured very precisely at the Brookhaven E821 experiment~\cite{Bennett:2006fi, Roberts:2010cj},
which is deviated from the SM prediction at the level more than 3$\sigma$~\cite{Hagiwara:2011af, Davier:2010nc}. 
In order to resolve the discrepancy, the additional contribution to $a_{\mu}$ of $\mathcal{O}(10^{-9})$ is required.
In the MSSM, if the smuons and chargino/neutralino are as light as $\mathcal{O}$(100)\,GeV for $\tan\beta=\mathcal{O}(10)$, 
the SUSY contribution to the muon $g-2$ is large enough and the anomaly is explained~\cite{Lopez:1993vi, Chattopadhyay:1995ae, Moroi:1995yh}. However, this clearly implies a tension: the observed Higgs boson mass suggests the heavy SUSY particles while the muon $g-2$ anomaly suggests the light SUSY particles, which arouses us to construct a non-trivial model. 

\vsp
In fact, there are ways suggested to resolve the tension:
\begin{itemize}
\item[(a)] New contributions to the Higgs boson mass: 
if there is an additional contribution to the Higgs boson mass, 
the SUSY particles are not necessarily heavy. 
In this case, the anomaly of the muon $g-2$ is explained by the contributions from the fairly light SUSY particles.
For instance, SUSY models with vector-like matter multiplets~\cite{Endo:2011mc, Moroi:2011aa, Endo:2011xq, Endo:2012cc}, 
the large trilinear coupling of the stop~\cite{Evans:2011bea, Evans:2012hg},
or an extra gauge interaction~\cite{Endo:2011gy} can accommodate both the observed Higgs boson mass and muon $g-2$ anomaly.\footnote{
In Ref.~\cite{Nakayama:2011iv}, it has been shown that 
in the next-minimal supersymmetric standard model,
the enhancement of the Higgs boson mass can be also applied to large $\tan\beta$ region 
once taking into account radiative corrections. This is favored by the muon $g-2$ explanation.
}

\item[(b)] Splitting masses for weakly interacting SUSY particles and strongly interacting ones:
with GUT breaking effects, 
it is possible to obtain light masses for the weakly interacting SUSY particles and heavy masses for 
the strongly interacting ones. This can be done for example in gauge mediation models with light colored 
and heavy non-colored messengers~\cite{Sato:2012bf, Ibe:2012qu, Bhattacharyya:2013xba, Bhattacharyya:2013xma},
or with the slepton multiplets embedded in extended SUSY multiplets at the messenger scale~\cite{Shimizu:2015ara}.

\item[(c)] Splitting masses for the first two and third generation sfermions:
instead of the splitting mass spectra for the strongly and weakly interacting SUSY particles, 
the sfermion masses can be split as in the case of SM fermions. With small masses for the first two generation sfermions of $\mathcal{O}(0.1 {\mathchar`-} 1)$\,TeV and the large masses for the third generation sfermions of $\mathcal{O}(10)$\,TeV, the tension between the Higgs boson mass of 125\,GeV and the muon $g-2$ anomaly is resolved~\cite{Ibe:2013oha}.

\end{itemize}

In this paper, we proposed a scenario with the splitting mass spectra corresponding to the case (c).\footnote{
See also e.g. Refs.~\cite{Mohanty:2013soa, Huh:2013hga, Iwamoto:2014ywa, Calibbi:2014yha, Harigaya:2015kfa, Kowalska:2015zja, Harigaya:2015jba, Chowdhury:2015rja, Belyaev:2016oxy, Okada:2016wlm} for other attempts to resolve the tension based on high energy models.
}
The splitting masses among the first two and third generation sfermions are naturally obtained by 
renormalization group (RG) running effects from Higgs soft masses~\cite{Yamaguchi:2016oqz}, if the squared values of the Higgs soft masses are negative. On the other hand, gaugino masses are generated by anomaly mediation~\cite{sequestered, Giudice:1998xp} with the gravitino mass of $\sim100$\,TeV. 
In our setup, the third generation sfermions have masses of $\mathcal{O}(10)$\,TeV, 
while the first/second generation sfermions have masses of $\mathcal{O}(0.1\,{\mathchar`-}\,1)$\,TeV 
without inducing the flavor changing neutral current (FCNC) problem.

The rest of the paper is organized as follows. In Sec. 2, we explain the setup of our model, based on the anomaly mediation. Only Higgs multiples couple to a SUSY breaking sector, which does not introduce the flavor changing neutral current problem. 
In Sec 3. we describe the mechanism for splitting mass spectra of first two generation and the third generation. 
Then, we show the consistent regions with the muon $g-2$ and the stop mass of $\mathcal{O}(10)$\,TeV. The unification of  Yukawa couplings is also discussed.
Finally, section 4 is devoted to the conclusion and discussions.

\section{Higgs-anomaly mediation}
We first explain the setup of our model. 
In our model, only the Higgs multiplets have direct couplings to a SUSY breaking field at the tree level. 
The other sparticle masses are generated radiatively via anomaly mediation effects and RG running effects from the Higgs soft masses.
The K{\"a}lher potential is given by
\begin{eqnarray}
K = - 3 M_P^2 \ln \left[1 - \frac{f(Z, Z^\dag) + \phi_i^\dag \phi_i + \Delta K}{3 M_P^2} \right], 
\end{eqnarray}
where $M_P$ is the reduced Planck mass, $Z$ is a SUSY breaking field, and $\phi_i$ is a MSSM chiral superfield. 
It is assumed that the vacuum expectation value (VEV) of $Z$ is much smaller than $M_P$. 
The SUSY is broken by the $F$-term of $Z$, $|\left<F_Z\right>| = \sqrt{3}\, m_{3/2}$, where $m_{3/2}$ is the gravitino mass. 
Here, $\Delta K$ contains direct couplings of the Higgs multiplets to the SUSY breaking field $Z$:
\begin{eqnarray}
\Delta K =  c_Z \frac{|Z|^2}{M_P^2} (|H_u|^2 + |H_d|^2),
\end{eqnarray}
where $H_u$ and $H_d$ are the up-type and down-type Higgs, respectively; $c_Z$ is a coefficient of $\mathcal{O}(0.01\,{\mathchar`-}\,0.1)$, which is taken as a free parameter but is assumed to be positive. 
For simplicity, we assume $H_u$ and $H_d$ have the common coupling to $Z$.\footnote{
If the soft masses of the up and down-type Higgs are not the same, 
the RG running may give non-negligible contributions to the soft mass parameters via $U(1)_Y$ gauge interactions.
} This may be justified if $H_u$ and $H_d$ are embedded into a same GUT multiplet of $SO(10)$.
In the case $\Delta K=0$, the above K{\"a}hler potential takes a sequestered form~\cite{sequestered}, i.e., the MSSM fields do not have direct couplings to $Z$ at the tree level. In this case a sfermion mass  is  
\begin{eqnarray}
\left. m_{\phi_i}^2 \right|_{\Delta K=0}  =  -\frac{1}{4} 
\Bigl(
\frac{\partial \gamma_{\phi_i}}{\partial g_a} \beta_{g_a} + \frac{\partial \gamma_{\phi_i}}{\partial y_{k}} \beta_{y_k}
\Bigr) m_{3/2}^2
\, ,
\end{eqnarray}
where $\gamma_{\phi_i}$ is an anomalous dimension defined by $\gamma_{\phi_i} \equiv (\partial \ln Z_{\phi_i}/ \partial \ln \mu_R)$; $g_a$ ($y_k$) is a gauge (Yukawa) coupling;
$\beta_{g_a}$ ($\beta_{y_k}$) is a beta-function of $g_a$ ($y_k$);
$\mu_R$ is a renormalization scale. 
Notice that the masses of the first and second generation sleptons are inevitably tachyonic, since the beta-functions for the  $SU(2)_L$ and $U(1)_Y$ gauge couplings are positive and Yukawa couplings are negligibly small~\cite{sequestered}. This problem is called the tachyonic slepton problem.

However, in our setup, the Higgs multiplets have soft masses of $\mathcal{O}(10)$\,TeV from $\Delta K$ for $m_{3/2} \sim 100$\,TeV, which play significant roles in low-energy SUSY mass spectra via the RG running: the tachyonic slepton problem is avoided and the masses of the third generation sfermions including the stops become $\mathcal{O}(10)$\,TeV 
if the Higgs soft mass squared is negative 
and $\tan\beta\, (\equiv \left<H_u^0\right>/ \left<H_d^0\right>)$ is large.

It is also assumed that there are no direct couplings between gauge field strength superfields and the SUSY breaking field,
which may originate from the fact that $Z$ is charged under a symmetry in the hidden sector.
Gaugino masses vanish at the tree level and are generated radiatively from anomaly mediation~\cite{sequestered, Giudice:1998xp}:
\begin{eqnarray}
M_1 &\simeq& \frac{33}{5} \frac{g_1^2}{16\pi^2} m_{3/2} \, , \ \ M_2 \simeq \frac{g_2^2}{16\pi^2} m_{3/2}, \ \ M_3 \simeq -3 \frac{g_3^2}{16\pi^2}  m_{3/2},
\end{eqnarray}
where $M_1$, $M_2$ and $M_3$ are the bino, wino and gluino, respectively; and
$g_1$, $g_2$ and $g_3$ are the gauge coupling constants of $U(1)_Y$, $SU(2)_L$ and $SU(3)_C$.\footnote{
The normalization of $g_1$ is taken to be consistent with $SU(5)$ GUT.
}
These masses are expected to be $\mathcal{O}(0.1 {\mathchar`-} 1)$\,TeV.

So far, the parameters of our model are summarized as
\begin{eqnarray}
m_{3/2}, \ m_{H}^2, \  \tan\beta, \ {\rm sign}(\mu),
\end{eqnarray}
where the boundary condition of the soft SUSY breaking parameters is set at $M_{\rm inp}=10^{16}$\,GeV $(\approx M_{\rm GUT})$;
$m_H^2 = m_{H_u}^2 (M_{\rm inp}) = m_{H_d}^2(M_{\rm inp})$, where $m_{H_u}$ and $m_{H_d}$ are the soft masses for the up-type Higgs and down-type Higgs, respectively.
We fix sign($\mu$) to be positive in the following discussions since we are interested in regions consistent with the muon $g-2$ experiment. 
In the parameter space of our interest, the typical values for $m_{3/2}$ and $m_H$ are $\sim 100$\,TeV and $\mathcal{O}(10)$\,TeV, respectively,  with $m_H^2<0$.

\section{Splitting mass spectra and the muon $g-2$}

Next, we explain how the mass hierarchy between the first/second and third generation sfermions are obtained.
As noted in \cite{Yamaguchi:2016oqz}, the hierarchical mass spectrum is realized when $m_{H_{u,d}}^2$ are negative and large.
Contributions from one-loop renormalization group equations (RGEs) raise the third generation sfermion masses via terms proportional to the squared of the Yukawa couplings: 
\begin{eqnarray}
\beta_{m_{Q_3}^2, m_{L_3}^2}&\ni& {1 \over 16 \pi^2} \left({  2y_t^2 m_{H_u}^2 + 2 y_b^2 m_{H_d}^2, 
~ 2 y_{\tau}^2 m_{H_d}^2}\right), \nonumber \\
\beta_{m_{\bar T}^2, m_{\bar B}^2, m_{\bar \tau}^2}&\ni& {1 \over 16 \pi^2} \left({  4 y_t^2 m_{H_u}^2, 4 y_b^2 m_{H_d}^2, 
~ 4 y_{\tau}^2 m_{H_d}^2}\right), \label{eq:rge_3rd}
 \end{eqnarray}
 where $Q_3$ and $L_3$ are the $SU(2)$ doublet squark and slepton of the third generation; 
 $\bar T$, $\bar B$ and $\bar \tau$ are the right-handed stop, sbottom and stau, respectively;
 $y_t$ is the top Yukawa coupling; and $m_{H_u}^2 \sim m_{H_d}^2$. 
 The contributions from Eq.~(\ref{eq:rge_3rd}) dominate those from anomaly mediation.
 After solving RGEs, the third generation sfermions obtain masses of $\sim 10$\,TeV at the low energy scale
 for $m_{H} \sim 10$\,TeV.

 In addition to the contributions from anomaly mediation, 
 the sfermions of the first two generations also obtain masses from $m_{H_u}^2$ and $m_{H_d}^2$ via the RG running, though they are suppressed, compared to those of the third generation sfermions. 
 This is because one-loop terms in RGEs are proportional
  to the squared of the first/second generation Yukawa couplings, and gauge interaction terms are at the two-loop level. 
As a consequence, the sfermion mass spectrum at the low energy scale becomes automatically hierarchical. 
Note that the tachyonic slepton masses are avoided 
due to negative terms involving $m_{H_{u,d}}^2 (<0)$ 
and terms involving the Yukawa coupling squares and $g_1^2$ in beta-functions at the two-loop level:
\begin{eqnarray}
\beta_{m_{L_i}^2} &\ni& \frac{1}{(16\pi^2)^2} \left[\left(3 g_2^4  + \frac{9}{25} g_1^4\right)( m_{H_u}^2 + m_{H_d}^2 )\,  -\frac{6}{5}g_1^2  S' \right] , \nonumber \\
\beta_{m_{\bar E_i}^2} &\ni& \frac{1}{(16\pi^2)^2} \left[\left( \frac{36}{25} g_1^4 \right)( m_{H_u}^2 + m_{H_d}^2 ) + \frac{12}{5} g_1^2 S' \right]\,,
\end{eqnarray}
where ${\bar E}_3=\bar \tau$ and 
\begin{eqnarray}
S' &\ni& - 3 m_{H_u}^2 y_t^2 + 3 m_{H_d}^2 y_b^2 + m_{H_d}^2 y_{\tau}^2.
\end{eqnarray}

\begin{figure}[!t]
\begin{center}
\includegraphics[scale=1.15]{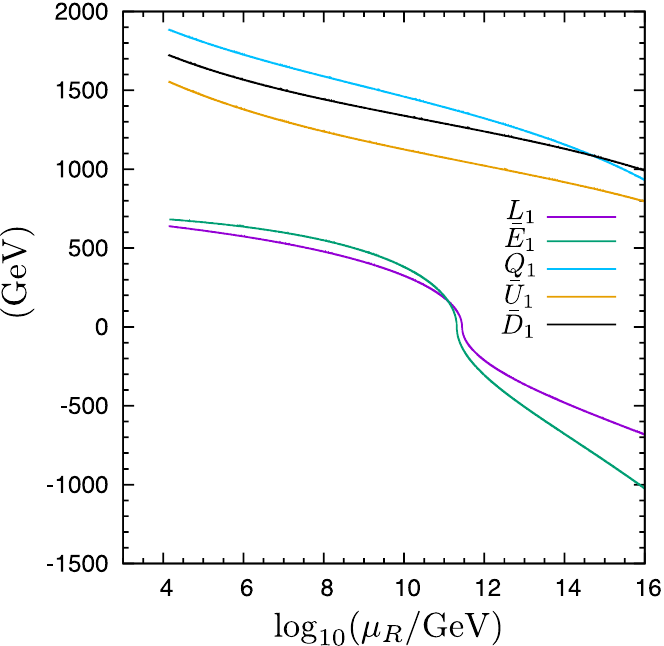}
\includegraphics[scale=1.15]{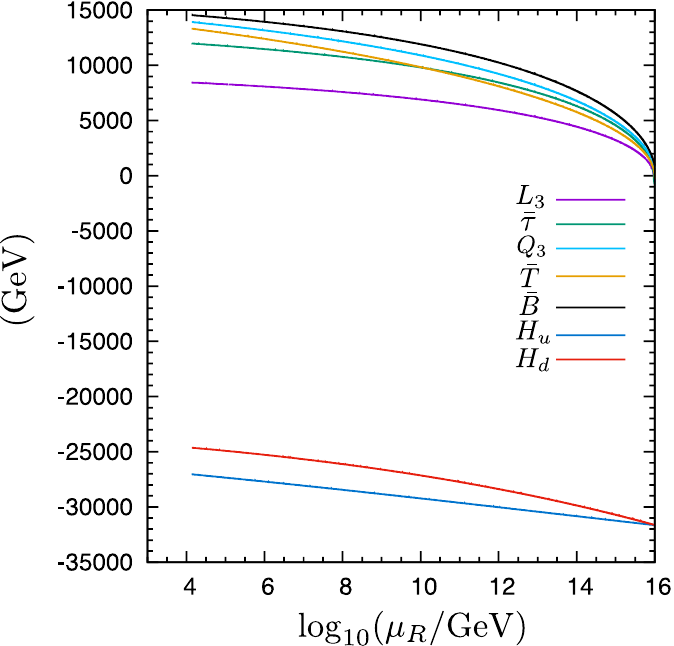}
\caption{
RG runnings of soft SUSY breaking masses as functions of the renormalization scale $\mu_R$.
The runnings of the first generation sfermion masses are shown in the left panel,
and those of the third generation sfermion masses and Higgs soft masses are shown in the right panel.
We take $m_{3/2}=120$\,TeV, $\tan\beta=48$, and $m_{H}=-10^9$\,GeV$^2$. Here, $\alpha_s(m_Z)=0.1185$ and $m_t({\rm pole})=173.34$\,GeV.
}
\label{fig:rge}
\end{center}
\end{figure}

In Fig.\,\ref{fig:rge}, we show the RG running of soft mass parameters for $m_{3/2}=120\,{\rm TeV}, m_H=-10^9\,{\rm GeV}^2$ and $\tan\beta=48$. Here and hereafter, we take $\alpha_s(m_Z) = 0.1185$~\cite{pdg} and $m_{t}({\rm pole})=173.34$\,GeV~\cite{topmass}. 
In the left panel, the runnings of the first generation sfermion masses are shown. 
The runnings of the third generation sfermion masses and the Higgs soft masses are shown in the right panel.
One can see that the masses of the first generation sfermions at the low energy scale are $\mathcal{O}(0.1$-1)\,TeV, 
avoiding the tachyonic masses for the sleptons ($L_1$ and $\bar E_1$). 
On the other hand, the masses for the third generation sfermions 
including the stop masses grow rapidly as $\mu_R$ decreases, and they reach to $\sim 10$\,TeV at the low energy scale.

\vsp

In our setup, for the successful electroweak symmetry breaking (EWSB) with $\tan\beta = \mathcal{O}(10)$, 
the $\mu$-term has to be large as $\mathcal{O}(10)$\,TeV and $m_{H_d}^2 - m_{H_u}^2 \gtrsim 0$, which is required to avoid the tachyonic mass for the CP-odd Higgs. 
The latter condition can be satisfied only when the bottom- and tau- 
Yukawa couplings, $y_b$ and $y_{\tau}$, are large enough, which enters beta-functions for $m_{H_u}^2$ and $m_{H_d}^2$ as 
\begin{eqnarray}
\beta_{m_{H_u}^2, m_{H_d}^2} \ni  {1 \over 16 \pi^2} \left({ 6y_t^2 m_{H_u}^2 ,~ 6y_b^2 m_{H_d}^2 +2y_{\tau}^2 
m_{H_d}^2}\right).
 \end{eqnarray} 
The absolute value of the negative $m_{H_d}^2$ decreases 
more than that of $m_{H_u}^2$ for large $y_b$ and $y_\tau$, if $|\beta_{m_{H_d}^2}|$ is larger than $|\beta_{m_{H_u}^2}|$. 
 In our model, this is achieved with a large $\tan \beta \sim 40-50$, 
 since the bottom Yukawa coupling gets larger with threshold corrections~\cite{Hall:1993gn, Carena:1994bv},  
 \begin{eqnarray}
 y_b={m_b  \over v_d ( 1 + \Delta_b)} ,  
 \ \  \Delta_b \ni  \frac{g_3^2 }{6\pi^2} \mu M_3 \tan\beta \, I(m_{\tilde b_1}^2, m_{\tilde b_2}^2, M_3^2), \label{eq:yb_cor}
 \end{eqnarray}
 where $m_b$ is a bottom quark mass;
  $v_d$ is a VEV of the down-type Higgs;
 $m_{\tilde b_1}$ ($m_{\tilde b_2}$) is the mass of the lighter (heavier) sbottom;
 $I(x,y,z)$ is a loop function,
 \begin{eqnarray}
I(x,y,z) = -\frac{xy\ln (x/y) + yz \ln (y/z) + zx \ln (z/x)}{(x-y)(y-z)(z-x)}.
\end{eqnarray}
  Notice that the bottom Yukawa is enhanced when $\mu M_3$ is negative for the fixed $\tan\beta$,
 while the SUSY contribution to the muon $g-2$ is positive when $\mu M_1$ is positive. 
 In our case, both $\mu M_3<0$ and $\mu M_1>0$ can be satisfied since the anomaly induced gaugino masses are proportional to the $\beta$ functions of the corresponding gauge couplings. 
Note that the threshold correction to $y_b$, $\Delta_b$, improves the unification of the Yukawa couplings.

\begin{figure}[!t]
\begin{center}
\includegraphics[scale=0.6]{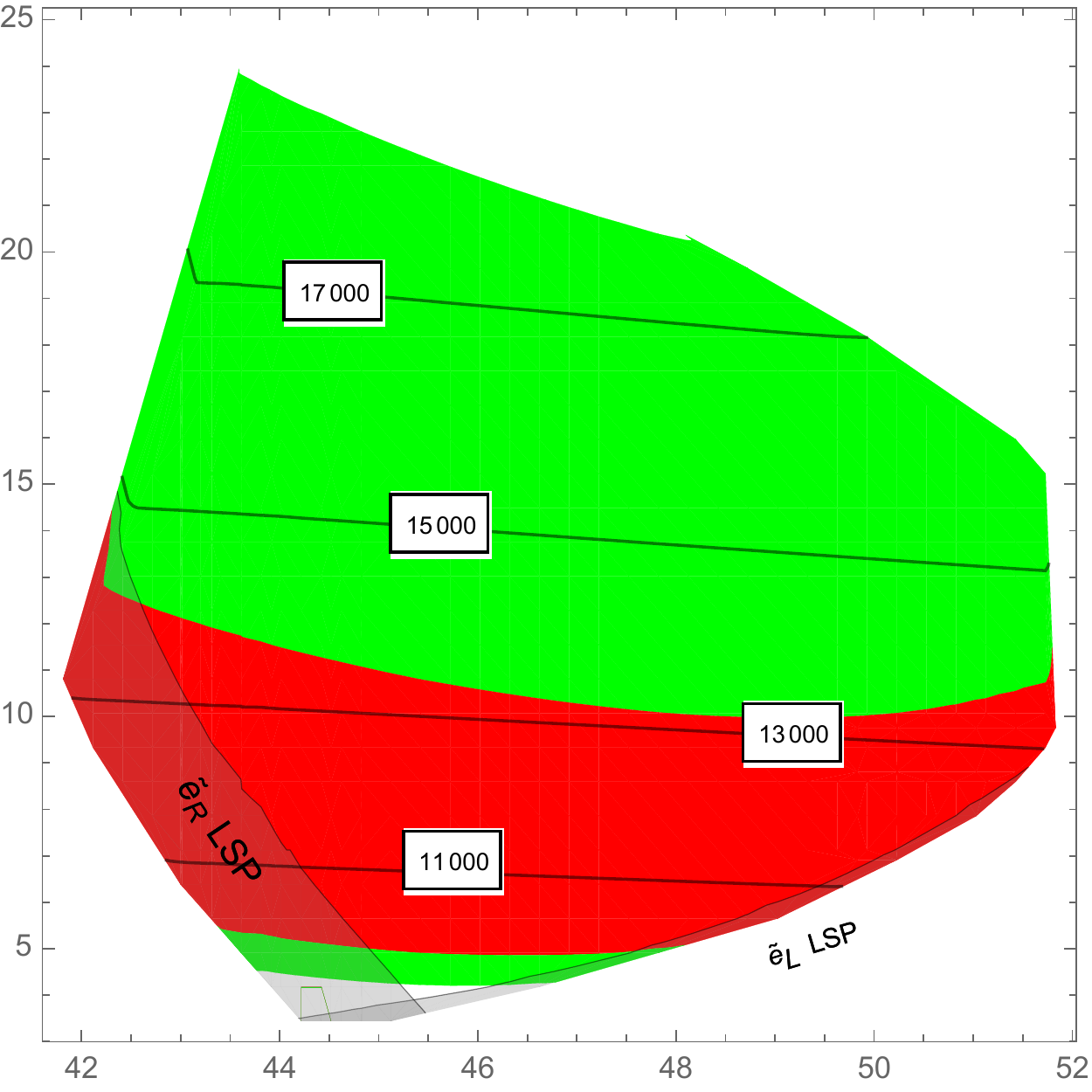}
\includegraphics[scale=0.6]{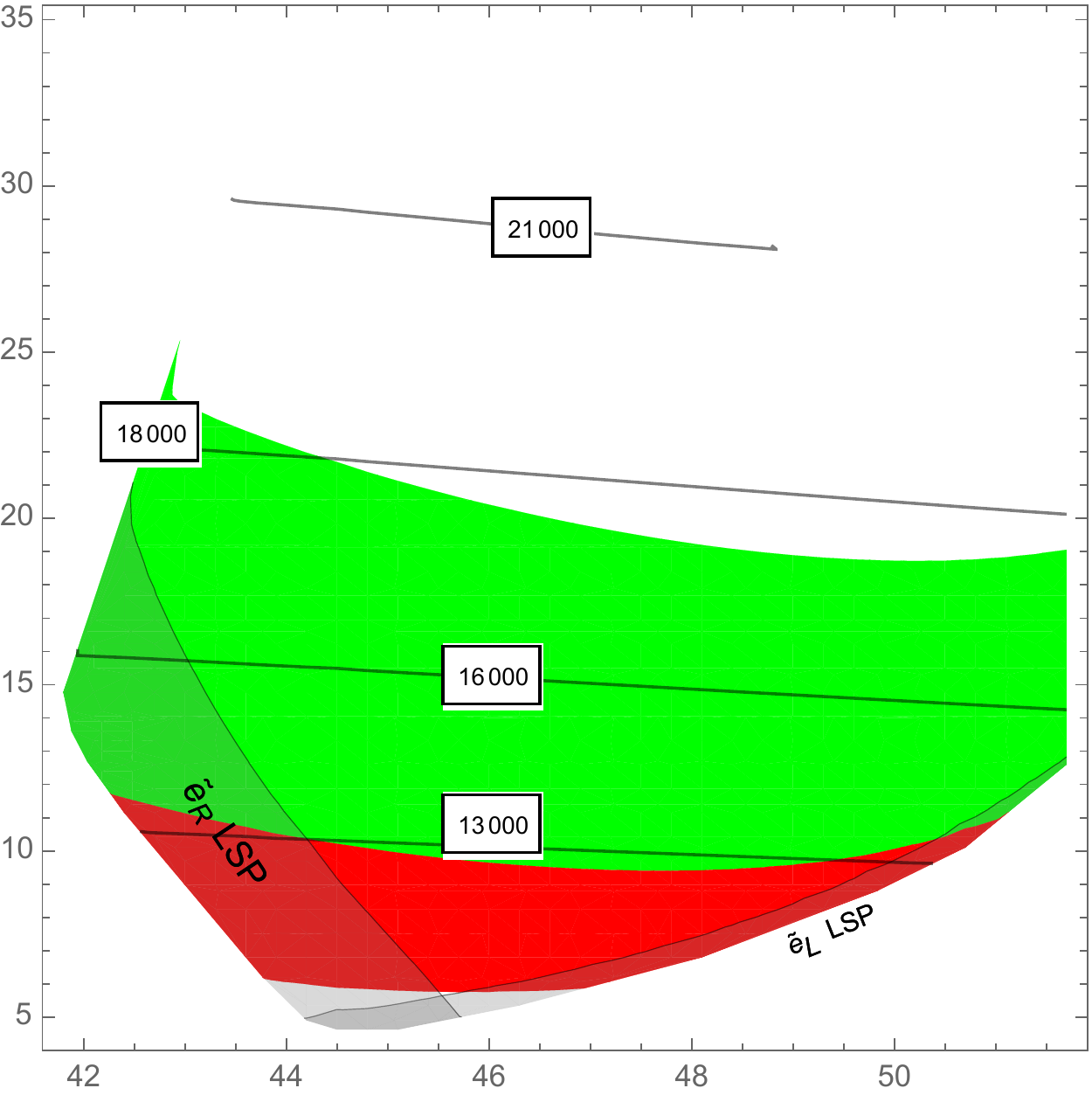}
\includegraphics[scale=0.6]{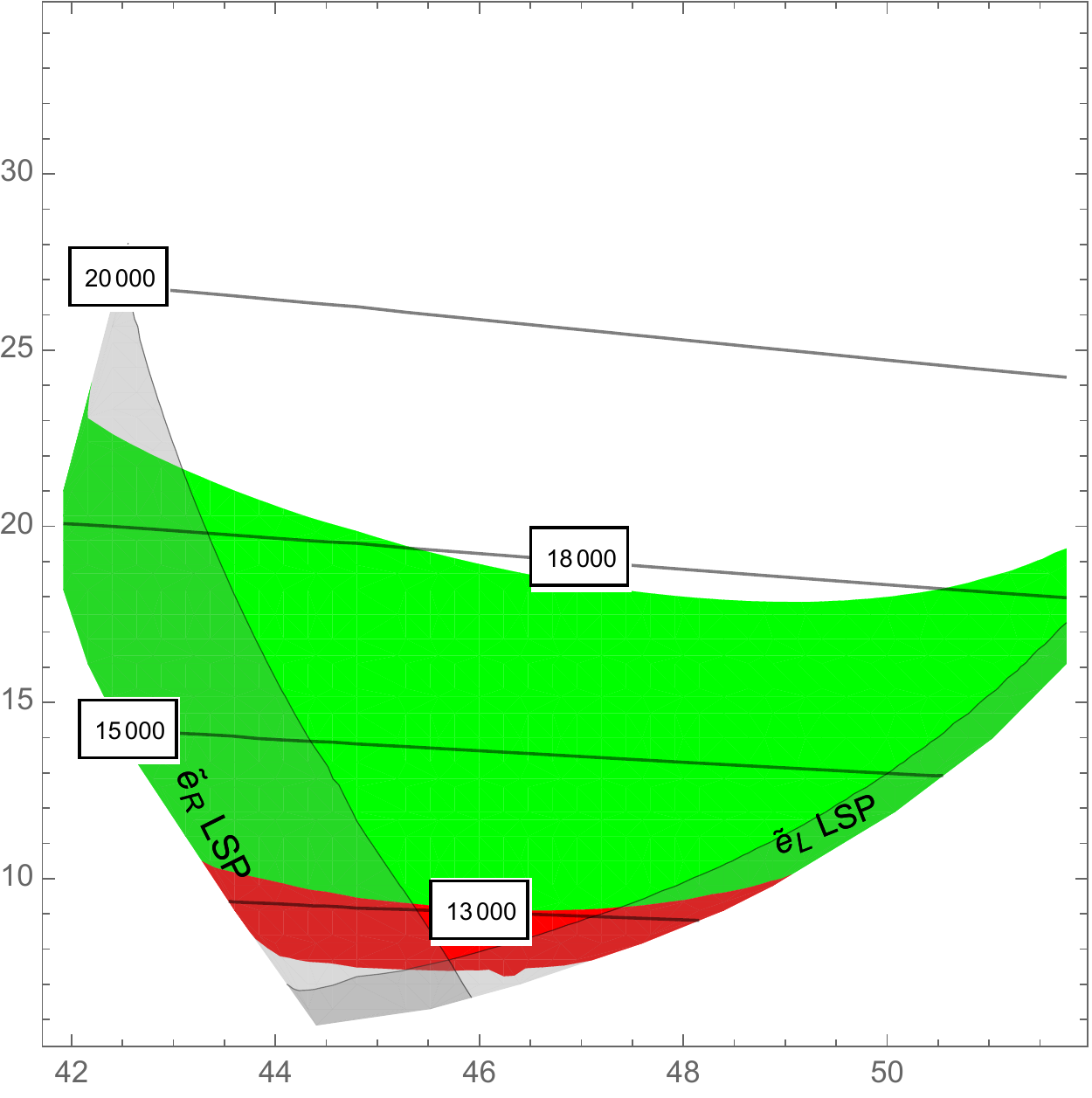}
\includegraphics[scale=0.6]{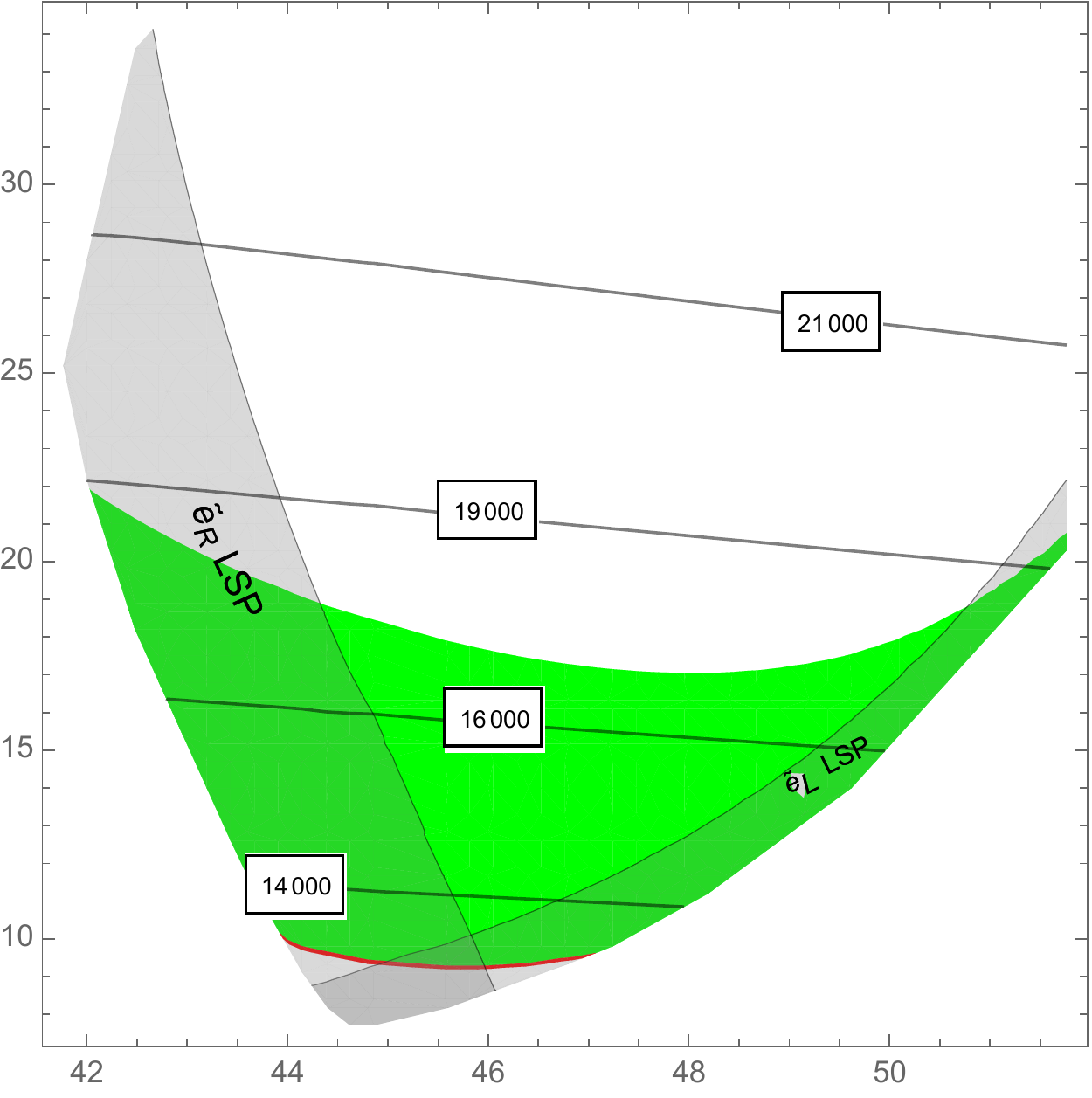}
\caption{
Contours of the stop mass in unit of GeV and $(a_{\mu})_{\rm SUSY}$ on $\tan\beta$-$c_H$ plane
for $m_{3/2}=$100 (top-left), 120 (top-right), 140 (bottom-left), 160 \,TeV (bottom-right). 
In the red (green) region, the muon $g-2$ is explained at 1$\sigma$ (2$\sigma$) level.
}
\label{fig:stop_gm2}
\end{center}
\end{figure}

\begin{figure}[!t]
\begin{center}
\includegraphics[scale=0.6]{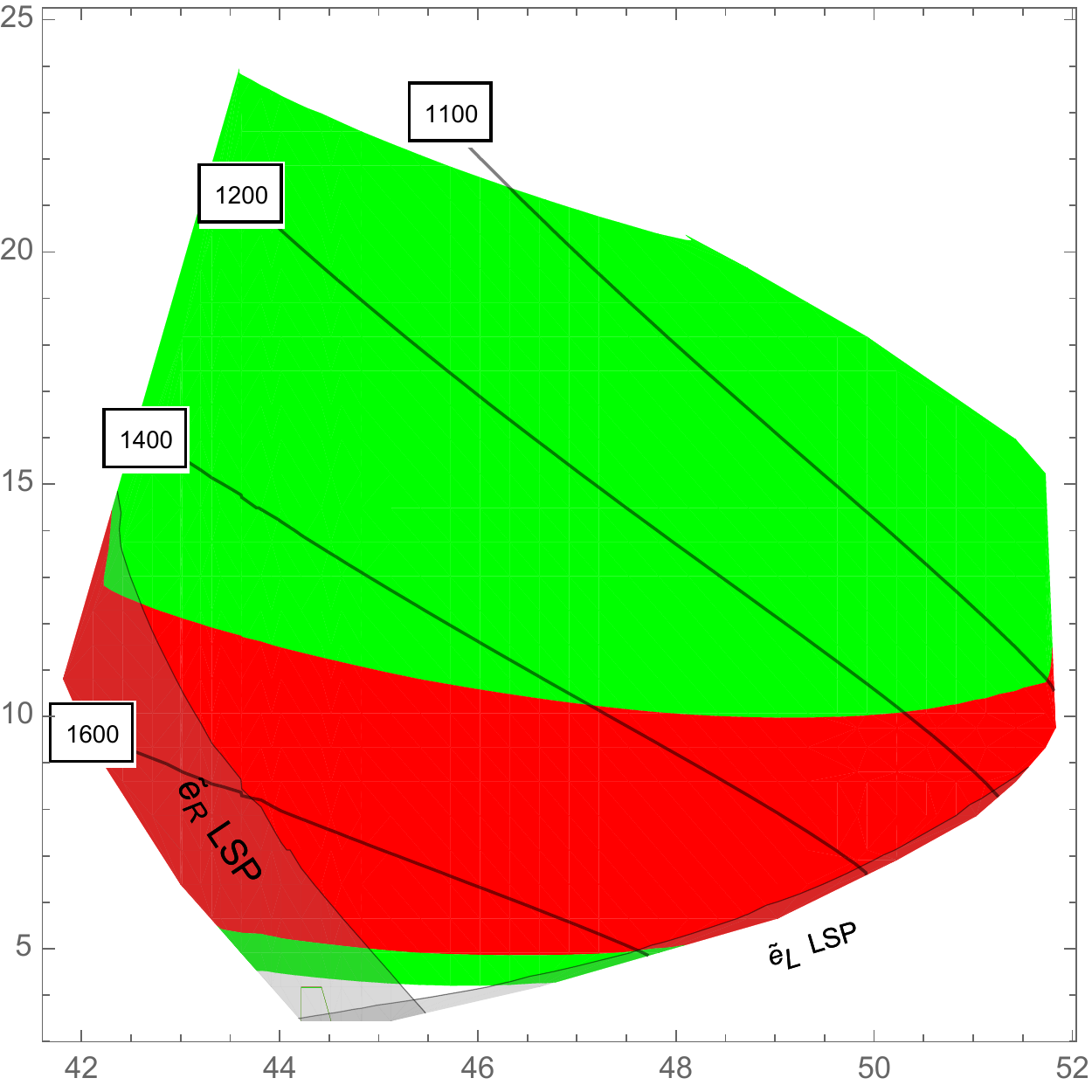}
\includegraphics[scale=0.6]{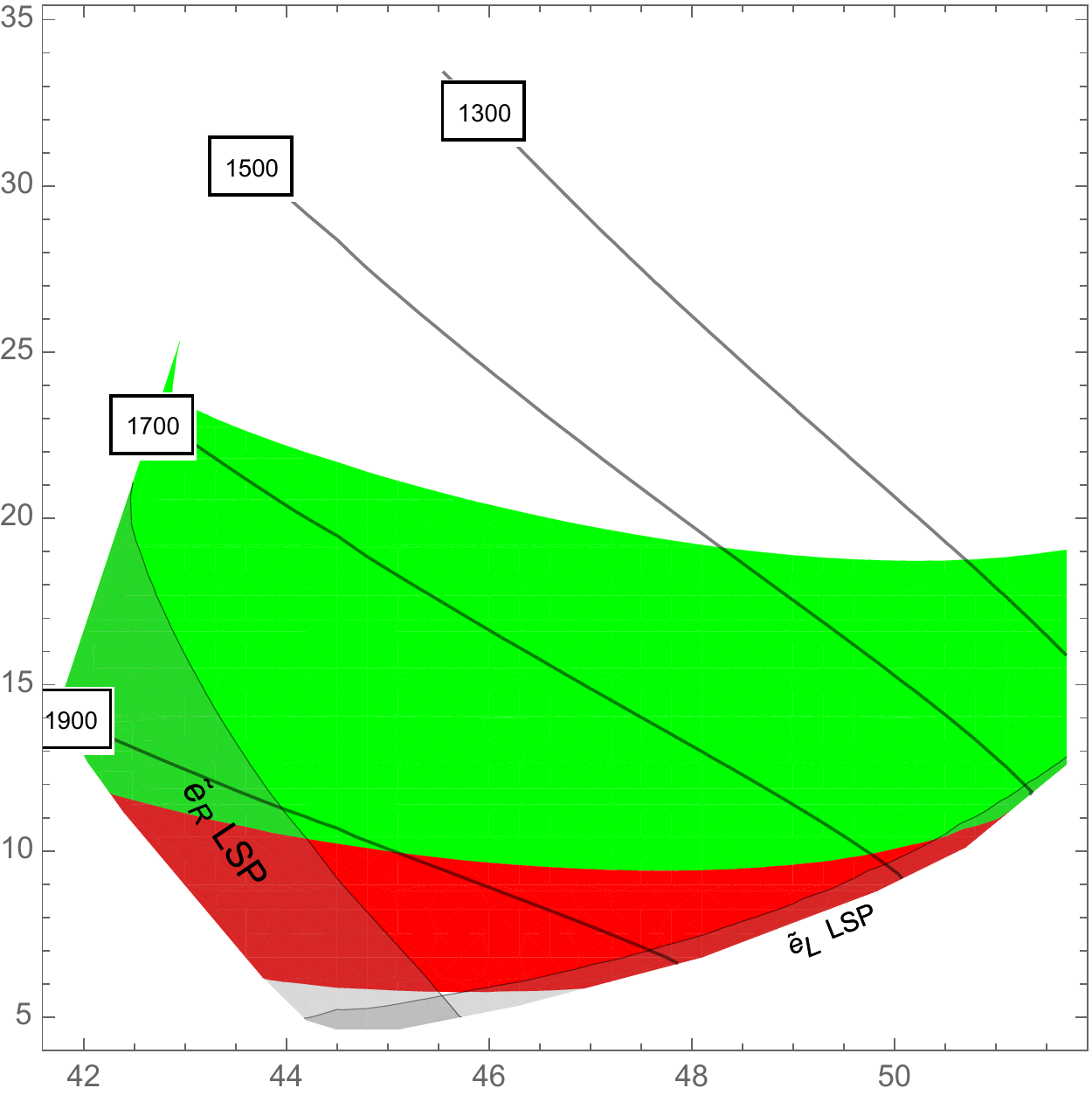}
\includegraphics[scale=0.6]{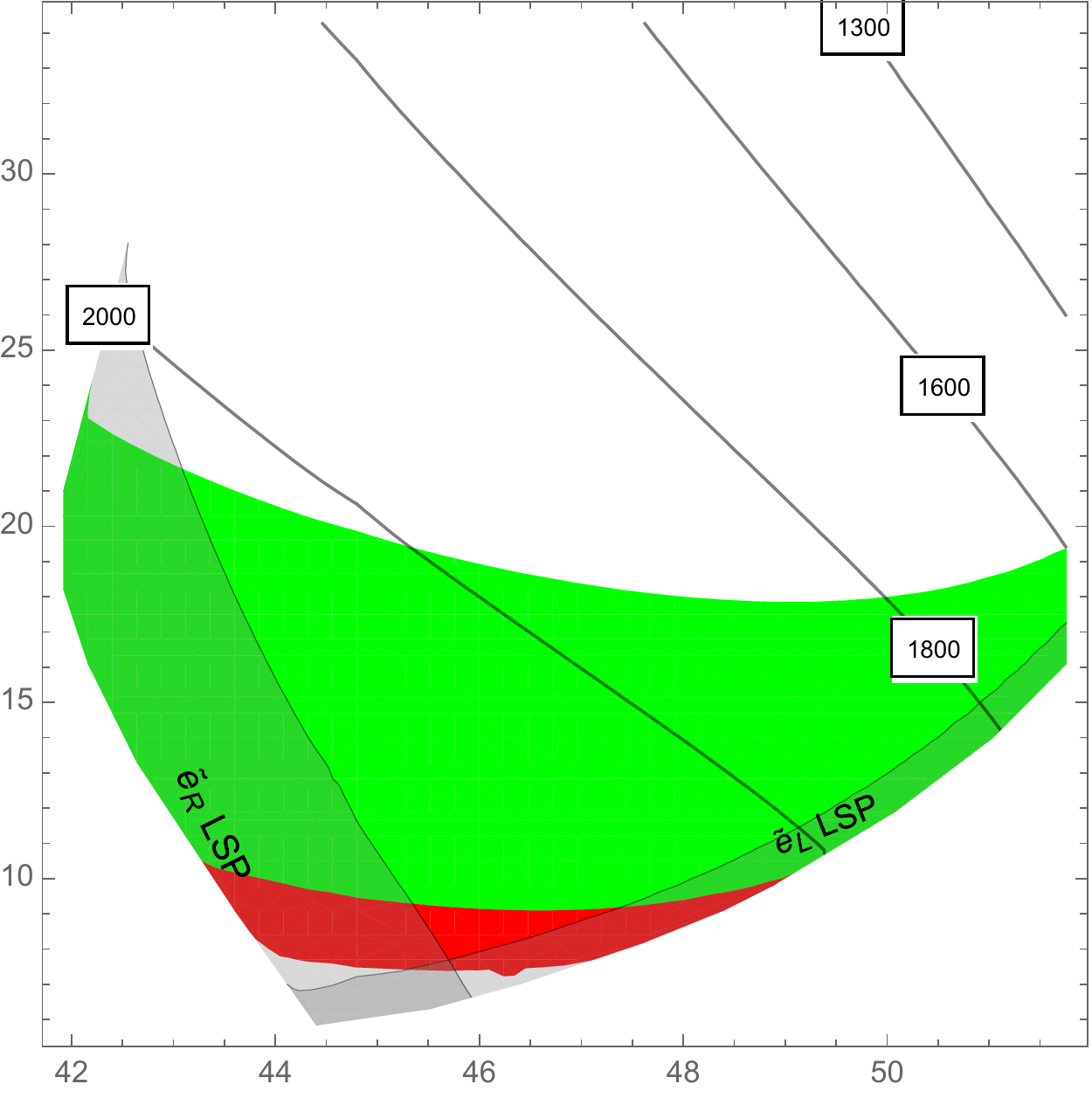}
\includegraphics[scale=0.6]{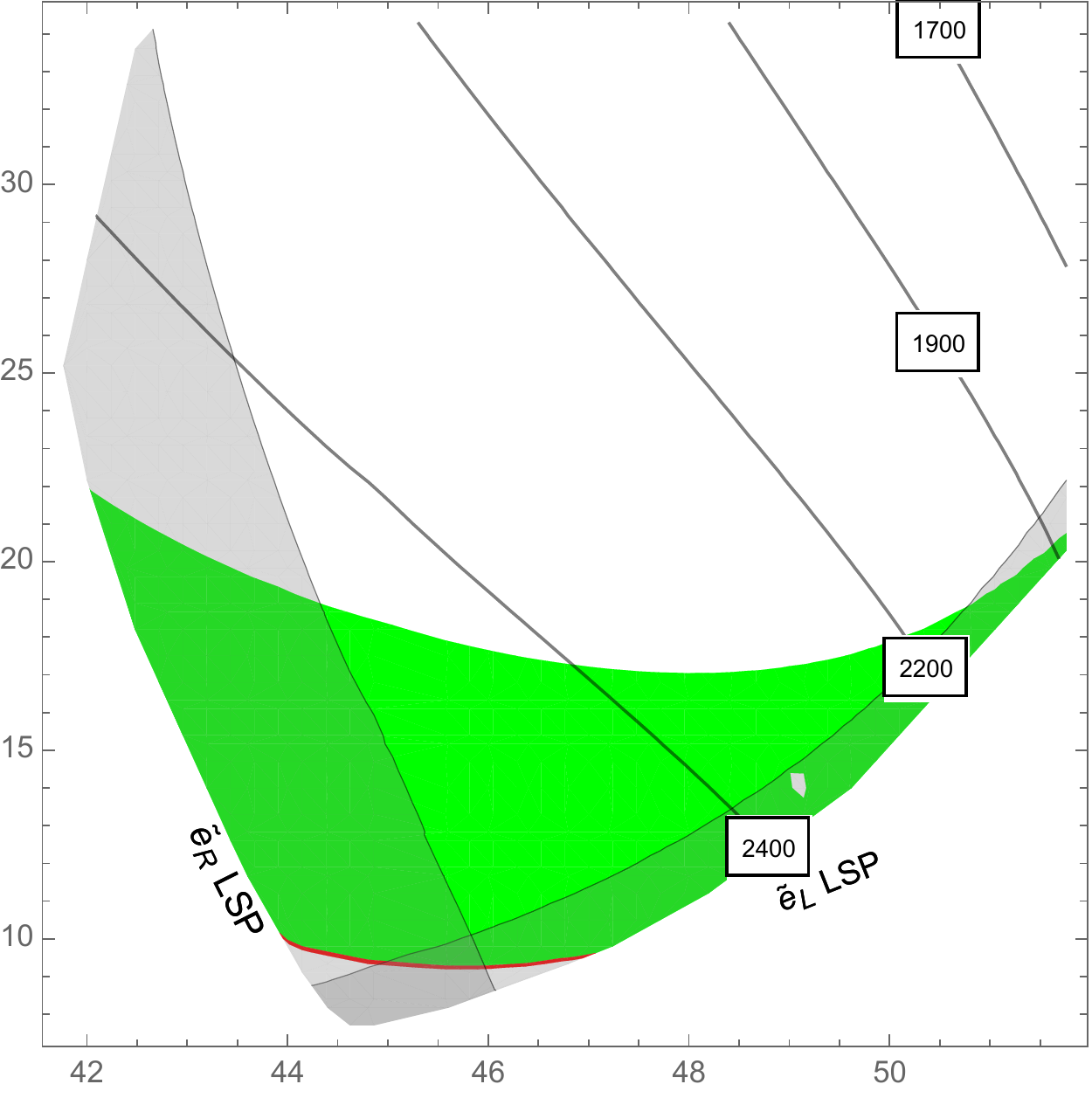}
\caption{
Contours of the up-type squark mass in unit of GeV on $\tan\beta$-$c_H$ plane for 
$m_{3/2}=$100 (top-left), 120 (top-right), 140 (bottom-left), 160\,TeV (bottom-right).
}
\label{fig:msq_gm2}
\end{center}
\end{figure}

\begin{figure}[!t]
\begin{center}
\includegraphics[scale=0.6]{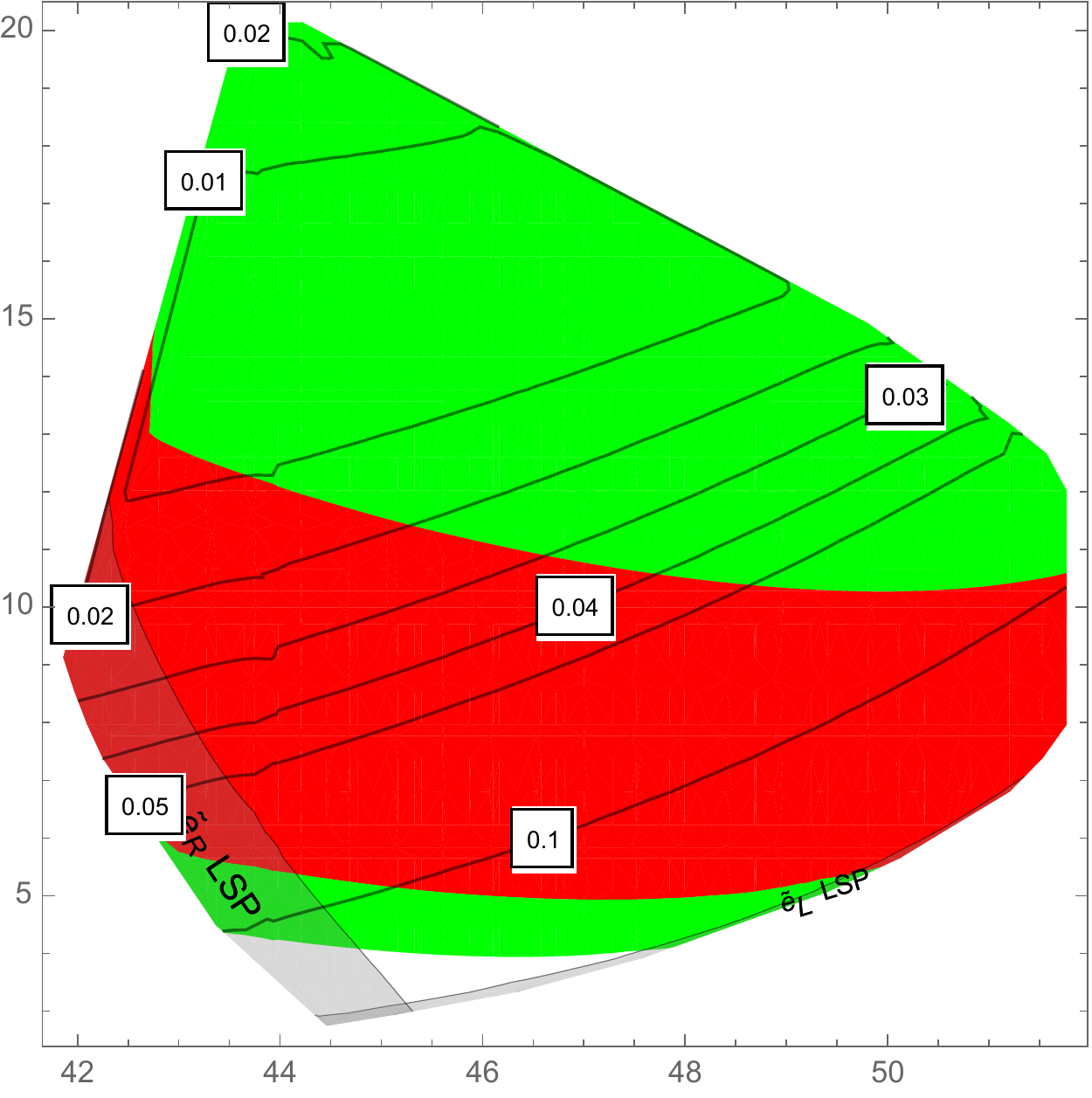}
\includegraphics[scale=0.6]{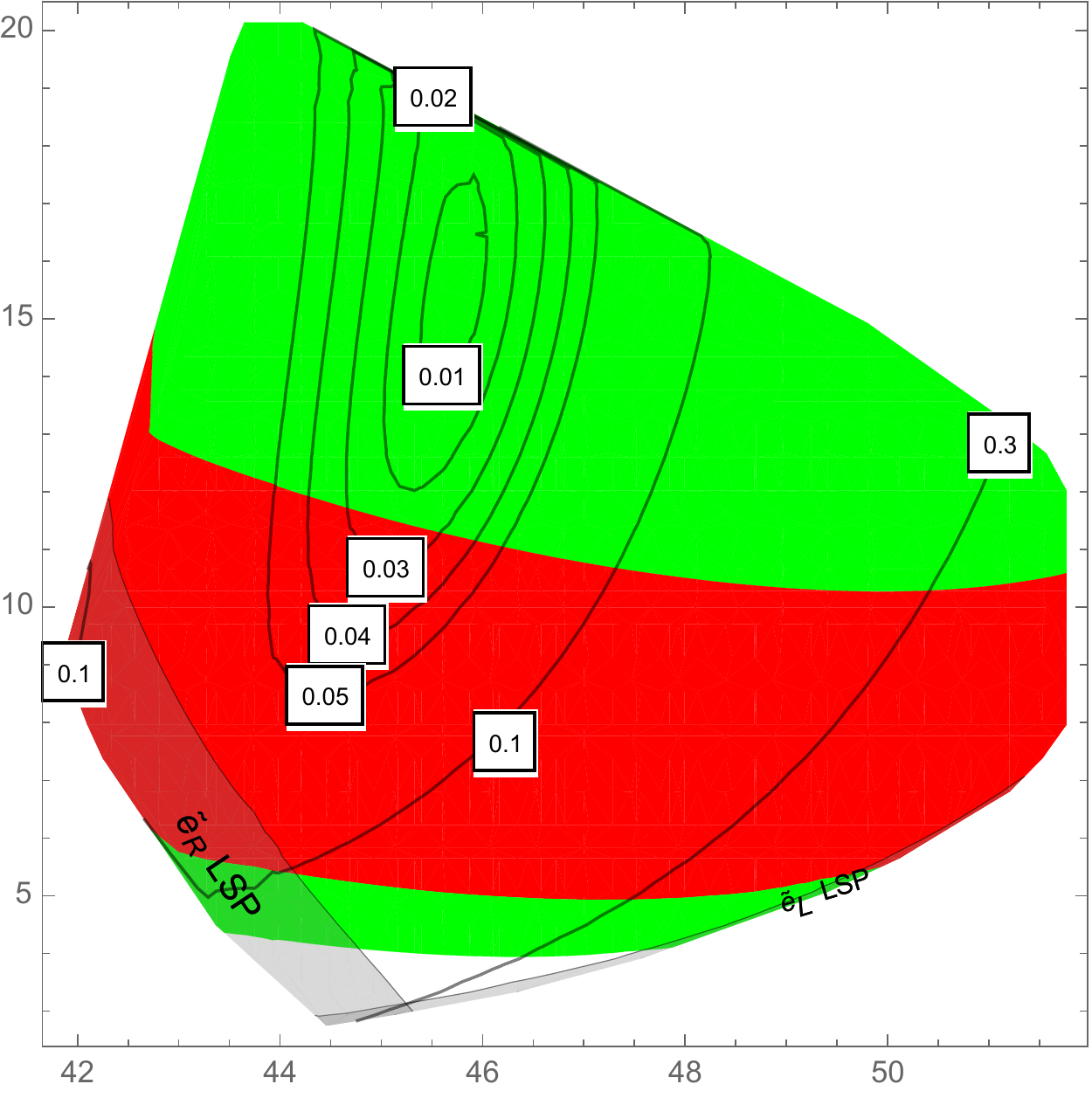}
\caption{
The unification of the Yukawa couplings is demonstrated on $\tan\beta$-$c_H$ plane. 
In the left panel (right panel), the contours of $\delta y_5$ ($\delta y_{10}$) are shown. We take $m_{3/2}=90$\,TeV.
}
\label{fig:yukawa_gut}
\end{center}
\end{figure}

\paragraph{Muon $g-2$}

In the typical parameter space of our model, the SUSY contribution to the muon $g-2$, $(a_\mu)_{\rm SUSY}$, is 
dominated by the bino-(L-smuon)-(R-smuon), where L and R represent the left-handed and right-handed, respectively. 
This contribution is given by~\cite{Moroi:1995yh}
\begin{eqnarray}
(\alpha_\mu)_{\rm SUSY} \simeq \left(
\frac{1 - \delta_{\rm QED}}{1 + \Delta_\mu }
\right) \frac{3}{5}
{g_1^2 \over 16\pi^2}{ m_\mu^2  \mu \tan\beta \, M_1 \over m_{\tilde{\mu}_L}^2 m_{\tilde{\mu}_R}^2}
\,f_N\left( 
\frac{m_{\tilde{\mu}_L}^2}{M_1^2},
\frac{m_{\tilde{\mu}_R}^2}{M_1^2}
\right) , \label{eq:gm2}
\end{eqnarray}
where $m_{\mu}$ is the muon mass; 
$m_{\tilde{\mu}_L}$ $(m_{\tilde{\mu}_R})$ is the mass of the L-smuon (R-smuon);
$f_N(x,y)$ is a loop function with $f_N(1,1)=1/6$. 
Here, $\Delta_\mu$ and $\delta_{\rm QED}$  are two-loop corrections: 
$\Delta_\mu$ is the correction to the muon Yukawa coupling~\cite{Marchetti:2008hw}, 
\begin{eqnarray}
\Delta_\mu \simeq \mu \tan\beta \frac{3}{5} \frac{g_1^2 M_1}{16\pi^2} I(M_1^2, m_{{\tilde \mu}_L}^2, m_{{\tilde \mu}_R}^2),
\end{eqnarray}
which is positive and can become as large as $\sim 1.0$, and $\delta_{\rm QED}= (4\alpha/\pi) \ln (m_{\tilde \mu}/m_\mu)$ is a leading logarithmic correction from QED~\cite{Degrassi:1998es}, where $\alpha$ is the fine-structure constant and $m_{\tilde \mu}$ is a smuon mass scale.
The SUSY contribution to $a_\mu$ is enhanced with the large $\mu \tan\beta$ and light smuons, which is the character of our model.

In Fig.~\ref{fig:stop_gm2}, we show the contours of the stop mass defined by $m_{\tilde t} \equiv \sqrt{m_{Q_3} m_{\bar T}}$ 
and the regions consistent with the muon $g-2$. The horizontal (vertical) axis shows $\tan\beta$ $(c_H)$, 
where $m_H^2 = -c_H \times 10^8\,{\rm GeV}^2$.
The SUSY mass spectra and $(a_\mu)_{\rm SUSY}$ are calculated using {\tt SuSpect\,2.43}~\cite{suspect} with modifications to include $\Delta_\mu$ in Eq.\,(\ref{eq:gm2}) 
and the effects of the muon Yukawa coupling on RG equations. In the red (green) regions, the muon $g-2$ is explained at 1$\sigma$ $(2\sigma)$ level. As a reference value, we quote~\cite{Hagiwara:2011af}
\begin{eqnarray}
(a_\mu)_{\rm EXP} - (a_\mu)_{\rm SM} = (26.1 \pm 8.0) \times 10^{-10},
\end{eqnarray}
where $(a_\mu)_{\rm EXP}$ is the experimental value~\cite{Bennett:2006fi,Roberts:2010cj} and $(a_\mu)_{\rm SM}$ is a SM prediction. 
In those regions, the stop mass is as large as 11\,-\,17\,TeV. In the shaded regions, the L-selectron ($\tilde e_L$) 
or R-selectron ($\tilde e_R$) is the lightest SUSY particle (LSP), and these regions are considered to be excluded. 
The regions with $m_{\tilde e_L} \lesssim 250$\,GeV, $m_{\tilde e_R} \lesssim 200$\,GeV~\footnote{
Here, the cuts, $m_{\tilde e_L} \lesssim 250$\,GeV and $m_{\tilde e_R} \lesssim 200$\,GeV, are chosen for the convenience of the numerical calculations.
} or the unsuccessful EWSB are dropped.

In Fig.~\ref{fig:msq_gm2}, we show the contours of the up-type squark mass ($m_{\bar U_1}$), which is the lightest squark in most of the parameter space.
The mass of the up-type squark lies in the range of 1000\,-\,2500\,GeV 
in the region consistent the muon $g-2$ at 1$\sigma$ level, depending on the gravitino mass.
The gaugino masses at the SUSY mass scale of $\sim10\,$TeV are 
\begin{eqnarray}
M_1 (m_{\tilde t}) &\simeq& (965, 1160, 1350, 1540) \, {\rm GeV} \, , \nonumber \\
M_2 (m_{\tilde t}) &\simeq& (301, 360, 419, 477)  \, {\rm GeV}  \, , \nonumber \\
M_3 (m_{\tilde t}) &\simeq& (-1800, -2150, -2490, -2820)  \, {\rm GeV}  \, ,
\end{eqnarray}
for $m_{3/2}=(100, 120, 140, 160)\,{\rm TeV}$. 
In most of the parameter space, the lightest SUSY particle (LSP) is the wino-like neutralino,
whose mass is almost degenerate with that of the lightest chargino due to large $\mu$-term of $\mathcal{O}(10)$\,TeV. The mass difference dominantly comes from W/Z boson loops, which tours out to be about $160\,{\rm MeV}$~\cite{Ibe:2012sx}.
This chargino is searched at the LHC using a disappearing-track, which leads to a constraint on the mass to be larger than 270\,GeV with the cross section estimated assuming the direct production~\cite{Aad:2013yna}.\footnote{
This wino-like neutralino is difficult to become a dominant component of dark matter in the parameter region of our interest, since the constraint from the indirect detection utilizing $\gamma$-ray is severe~\cite{Hayashi:2016kcy}.
}
In the regions where the L/R selectron is the LSP, 
even if the selectron is unstable with an $R$-parity violation, 
a LHC constraint of multi-lepton final states~\cite{Aad:2015eda} is severe; therefore, these regions are probably excluded.


\paragraph{Yukawa unification}

In our model, $y_b$ and $y_{\tau}$ are nearly degenerated at $M_{\rm inp}=10^{16}$\,GeV
 in some regions of the parameter space. 
Furthermore, one can find a region where even the three Yukawa couplings, $y_b$, $y_{\tau}$ and $y_t$, are nearly degenerated.
The unification of the Yukawa couplings at $M_{\rm inp}$ is demonstrated in Fig.~\ref{fig:yukawa_gut}. 
Motivated by the $SU(5)$ GUT, the contours of $\delta y_5 = \sqrt{(y_b-y_{\tau})^2}$ is shown in the left panel, 
while $\delta y_{10} = \sqrt{(y_b-y_{\tau})^2 + (y_t-y_{\tau})^2 + (y_t-y_b)^2 }$ in the right panel motivated by the $SO(10)$ GUT.
Here, $\delta y_{5}$ and $\delta y_{10}$ are evaluated at $M_{\rm inp}$. 
The unification can be achieved at $\mathcal{O}(1)$\% level with the help of $\Delta_b$ in Eq.~(\ref{eq:yb_cor}).

\begin{table*}[!t]
\caption{Mass spectra in sample points. 
}
\label{tab:sample}
\begin{center}
\begin{tabular}{|c||c|c|c|c|}
\hline
Parameters & Point {\bf I} & Point {\bf II}  & Point {\bf III} & Point {\bf IV}\\
\hline
$m_{3/2} $ (TeV) & 120  & 140  & 98 & 150\\
$m_{H}^2$ (GeV$^2$)  & $-9 \times 10^8$  & $-9 \times 10^8$  & $-8 \times 10^8$ & $-9.5 \times 10^8$\\
$\tan\beta$  & 48  & 46.7  &  48.2 & 46.5\\
\hline
%
Particles & Mass (GeV) & Mass (GeV)& Mass (GeV) & Mass (GeV)\\
\hline
$\tilde{g}$ & 2550 & 2930 &2120  & 3120\\
$\tilde{q}$ & 1830\,-\,2110 & 2240\,-\,2470 & 1440\,-\,1730 & 2420\,-\,2640\\\
$\tilde{t}_{2,1}$ (TeV) & 13.1, 12.5 & 13.1, 12.6 & 12.1, 11.7 & 13.5, 12.9\\
$\tilde{b}_{2,1}$ (TeV) & 14.2, 13.4 & 14.2, 13.5 & 13.0, 12.4 & 14.6, 13.8\\
$\tilde \chi_1^0$/$\tilde{\chi}_{1}^\pm$ & 378 & 440  & 311 & 470\\
$\tilde{\chi}_2^0$ & 1100 & 1290 & 896 & 1380\\
$\tilde{e}_{L, R}$ & 549, 682 & 485, 586  & 619, 630& 481, 558\\
$\tilde{\mu}_{L, R}$ & 609, 778 & 544, 680  & 671, 729& 539, 657\\
$\tilde{\tau}_{2,1}$ (TeV) & 11.4,  8.0&  11.1, 7.8 & 10.8, 7.6& 11.3, 7.9\\
$H^\pm$\,(TeV) & 10.9 & 10.7 & 9.7&  11.2\\
$h_{\rm SM\mathchar`-like}$ & 127.3 &  125.1  & 125.1 & 125.0\\
\hline
$\mu$ (TeV) & 25.8  & 25.8  & 24.3 & 26.5\\
$(a_{\mu})_{\rm SUSY}$\,$(10^{-10})$ & 18.6 &  18.1 & 21.8 & 17.2\\
\hline
\end{tabular}
\end{center}
\end{table*}

\paragraph{FCNC}
One might worry about the flavor violating sfermion masses induced by the Yukawa couplings and large $m_{H_{u,d}}^2$. 
In fact, the generated flavor violating masses are not so large but not negligibly small. 
Using the leading log approximation, an off-diagonal element of the sfermion mass matrix is estimated as
\begin{eqnarray}
\Delta (\delta_{LL}^d)_{12} \simeq \frac{1}{8\pi^2} V_{td}^* V_{ts} Y_t^2 \frac{m_{H_u}^2}{m_{\tilde q}^2} \ln \frac{m_{\tilde q}}{M_{\rm inp}},
\end{eqnarray}
in the super-CKM basis with ${\rm Re}(V_{td}^* V_{ts}) \approx -3.4 \times 10^{-4}$. 
Here, $m_{\tilde q}$ is a typical squark mass. 
Thus, 
\begin{eqnarray}
{\rm Re} [ \Delta (\delta_{LL}^d)_{12} ] \approx
-0.015 \left( \frac{2 \, {\rm TeV}}{m_{\tilde q}} \right)^2 
\left( \frac{-m_{H_u}^2}{10^9 {\rm GeV}^2} \right),
\end{eqnarray}
which is consistent with the constraint from $\Delta M_K$~\cite{Gabbiani:1996hi}. 

\paragraph{Mass spectra}

Finally we show some mass spectra in our model parameter space (Table~\ref{tab:sample}), 
where $\tilde \chi_1^0$ (mass eigenstate) is the wino-like neutralino, $\tilde \chi_2^0$  is the bino-like neutralino and $\tilde g$ is the gluino.
The Higgs boson mass is computed using {\tt FeynHiggs\,2.12.0}~\cite{feynhiggs, feynhiggs2, feynhiggs3, feynhiggs4, feynhiggs5}. In these points, the stop mass is large as 12-13\,TeV 
while the first/second generation sfermions and gauginos are light as $\mathcal{O}$(0.1-1)\,TeV. 
The higgsino mass parameter, $\mu$, is as large as $\sim 20$\,TeV, leading to the fine-tuning of the EWSB scale as $(125\,{\rm GeV})^2/(2\mu^2) \sim 10^{-5}$. 
With the smuons of $\mathcal{O}(100)$ GeV and large $\tan\beta$ of $\sim 50$, the muon $g-2$ is explained at the 1$\sigma$ level.

\section{Conclusion and discussion}

We have proposed a scenario where only the Higgs multiplets have direct couplings to the SUSY breaking sector.
The standard model matter multiplets as well as the gauge multiples do not have direct couplings to the SUSY breaking field at the classical level, and their masses are generated radiatively by anomaly mediation and Higgs loops. 
Due to RG running effects from the Higgs soft masses of $\mathcal{O}$(10)\,TeV, 
the third generation sfermions have masses of $\mathcal{O}$(10)\,TeV while the first and second generation sfermions 
have masses of $\mathcal{O}$(0.1\,-\,1)\,TeV, avoiding the tachyonic slepton problem of anomaly mediation. 
The hierarchy of the masses originates from the structure of the Yukawa couplings, i.e., 
the Yukawa couplings of the third generation are much larger than those of the first and second generations. 
In this case, there is no SUSY FCNC problem. 
The hierarchical mass spectrum allows us to explain the Higgs boson mass of $\sim$\,125\,GeV and the observed value of the muon $g-2$, simultaneously. 
In the whole region explaining the muon $g-2$ anomaly, the masses of the light squarks and gluino lies in the range 
smaller than 3\,TeV; therefore, it is expected to be checked at the LHC Run-2 or the high luminosity LHC.

Since the gravitino is heavier than about 100\,TeV, the cosmological gravitino problem is relaxed~\cite{Kawasaki:2008qe}. 
Moreover, in our setup, the SUSY breaking field is not necessarily a gauge singlet of a hidden sector symmetry; therefore, the cosmological moduli problem or Polonyi problem~\cite{Coughlan:1983ci, Ellis:1986zt, Goncharov:1984qm, Banks:1993en, deCarlos:1993wie} can be avoided. 

The possible drawback of our setup is the origin of the Higgs $B$-term of $\mathcal{O}(1)$\,TeV with the $\mu$-term of $\mathcal{O}(10)$\,TeV. 
This may be due to the fine-tuning of an ultraviolet model. 
Alternatively, the $\mu$-term and $B_\mu$-term may be generated 
by the vacuum expectation values of a $A$-term and $F$-term of a singlet chiral superfield {\it a la}  the next-to minimal supersymmetric standard model.

\section*{Acknowledgments}
We would like to thank Yutaro Shoji for collaboration at an early stage of this work. 
This work is supported by JSPS KAKENHI Grant Numbers JP15H05889 (N.Y.) and JP15K21733 (N.Y.).

\end{document}